# A multi-platform metabolomics approach identifies novel biomarkers associated with bacterial diversity in the human vagina

Amy McMillan (1,2), Stephen Rulisa (3), Mark Sumarah (4), Jean M. Macklaim (1,5), Justin Renaud (4), Jordan Bisanz (1,2), Gregory B. Gloor (5), and Gregor Reid (1,2,6) *

1.Canadian Centre for Human Microbiome and Probiotic Research, Lawson Health Research Institute, Western University, London, Ontario, Canada
2.Department of Microbiology and Immunology, Western University, London, Ontario, Canada
3.University of Rwanda, and University Teaching Hospital of Kigali, Kigali, Rwanda
4.Agriculture and Agri-food Canada, London, Ontario, Canada
5.Department of Biochemistry, Western University, London, Ontario, Canada
6.Department of Surgery, Western University, London, Ontario, Canada

**Corresponding Author Information:**
Gregor Reid, Canadian R&D Centre for Probiotics,
F3-106, Lawson Health Research Institute,
268 Grosvenor Street, London, Ontario
N6A 4V2, Canada
Tel: 519-646-6100 x65256
gregor@uwo.ca

**Abstract**

Bacterial vaginosis (BV) increases transmission of HIV, enhances the risk of preterm labour, and its associated malodour impacts the quality of life for many women. Clinical diagnosis primarily relies on microscopy to presumptively detect a loss of lactobacilli and acquisition of anaerobes. This diagnostic does not reflect the microbiota composition accurately as lactobacilli can assume different morphotypes, and assigning BV associated morphotypes to specific organisms is challenging. Using an untargeted metabolomics approach we identify novel biomarkers for BV in a cohort of 131 Rwandan women, and demonstrate that metabolic products in the vagina are strongly associated with bacterial diversity. Metabolites associated with high diversity and clinical BV include 2-hydroxyisovalerate and γ-hydroxybutyrate (GHB), but not the anaerobic end-product succinate. Low diversity, and high relative abundance of lactobacilli, is characterized by lactate and amino acids. Biomarkers associated with diversity and BV are independent of pregnancy status, and were validated in a blinded replication cohort from Tanzania (n=45), in which we predicted clinical BV with 91% accuracy. Correlations between the metabolome and microbiota identified *Gardnerella vaginalis* as a putative producer of GHB, and we demonstrate production by this species *in vitro*. This work provides a deeper understanding of the relationship between the vaginal microbiota and biomarkers of vaginal health and dysbiosis.

## Significance statement

Bacterial vaginosis (BV) is the most common vaginal condition, characterized by an increase in bacterial diversity with a corresponding decrease in *Lactobacillus* species. Clinical diagnosis often relies on microscopy, which may not reflect the microbiota composition accurately. Here we identify novel biomarkers for BV, and demonstrate that the vaginal metabolome is strongly correlated with bacterial diversity. Metabolites associated with high diversity and clinical BV are common to both pregnant and non-pregnant women, and were replicated in a blinded cohort with high sensitivity and specificity. We pinpoint the organism responsible for producing one of these biomarkers, and demonstrate production by this species *in vitro*. This work provides novel insight into the metabolism of the vaginal microbiota and provides a foundation for improved detection of disease.

## Introduction

The vaginal microbiota is dominated by *Lactobacillus* species in most women, predominately by *L. iners* and *L. crispatus* (1-3). When these lactobacilli are displaced by a group of mixed anaerobes, belonging to the genus *Gardnerella, Prevotella, Atopobium* and others, this increase in bacterial diversity can lead to bacterial vaginosis (BV) (1-3). BV is the most common vaginal condition, affecting an estimated 30% of women at any given time (4). While many women remain asymptomatic (2-5), when signs and symptoms do arise they include an elevated vaginal pH>4.5, discharge, and malodor due to amines (6-8). BV is also associated with a number of comorbidities, including increased transmission and acquisition of HIV and other sexually transmitted infections (9), and increased risk of preterm labour (10).

In most instances, diagnosis is dependant upon microscopy of vaginal fluid to identify BV-like bacteria alone (Nugent Scoring (11)), or in combination with clinical signs (Amsel Criteria (12)). The precision and accuracy of these methods are poor due to the diverse morphology of vaginal bacteria, the observation that many women with BV are asymptomatic, and subjectivity in microscopic examination (13-15). Misdiagnosis creates stress for the patient, delays appropriate intervention and places a financial burden on the health care system. A rapid test based on stable, specific biomarkers for BV would improve diagnostic accuracy and speed, and reduce costs through improved patient management. .

Metabolomics, defined as the complete set of small molecules in a given environment, has been utilized in a variety of systems to identify biomarkers of disease (16,17), and provide functional insight into shifts in microbial communities (18). Using an untargeted multiplatform metabolomics approach, combined with 16S rRNA gene sequencing, we demonstrate that the vaginal metabolome is driven by bacterial diversity, and identify novel biomarkers of clinical BV that can be reproduced in a blinded validation cohort. We further demonstrate that *Gardnerella vaginalis*, which has long been thought to be an important contributor to BV is the likely source of one of the most specific compounds.

This work illustrates how changes in community structure alter the chemical composition of the vagina, and identifies highly specific biomarkers for a common condition.

## Results

### The vaginal metabolome is most correlated with bacterial diversity

We completed a comprehensive untargeted metabolomic analysis of vaginal fluid in two cross-sectional cohorts of Rwandan women: pregnant (P, n=67) and non-pregnant (NP, n=64) (**Table S1**). To normalize the amount of sample collected, vaginal swabs were weighed prior to and after collection and normalized to equivalent concentrations. This enabled us to collect precise measurements of metabolites in vaginal fluid. Metabolite profiling was carried out using both gas chromatography-mass spectrometry (GC-MS) and liquid chromatography-mass spectrometry (LC-MS), and microbiota composition by 16S rRNA gene sequencing.

The metabolome determined by GC-MS contained 128 metabolites (**Table S2**). We conducted a series of partial least squares (PLS) regression analyses to determine the single variable that could best explain the variation in the metabolome. In both cohorts, the diversity of the microbiota, as measured using Shannon's Diversity (19), was the factor that explained the largest percent variation in the metabolome (**Table S3**), demonstrating that the vaginal metabolome is most correlated with bacterial diversity (**Fig. 1A**). Metabolites robustly associated with this diversity (95% CI <> 0)(**Fig. 1B**) were determined by jackknifing, and within this group, metabolites associated with extreme diversity tended to have less variation in the jackknife replicates, and were common to both pregnant and non-pregnant women. This identified a core set of metabolites associated with diversity.

The two cohorts overlapped by principal component analysis (PCA) (**Fig S1**), and no metabolites were significantly different between pregnant and non-pregnant women (unpaired t-test, Benjamini-Hochberg $p > 0.01$). Thus, the cohorts were combined for all further analysis.

**Metabolites and taxa associated with diversity**

A single PLS regression was performed on all samples with Shannon's diversity as a continuous latent variable (**Fig S2**). Samples were then ordered by their position on the 1$^{st}$ component of this PLS. The diversity indices, microbiota and metabolites associated with diversity of PLS ordered samples are shown in **Fig. 2**. The vaginal microbiota of Rwandan women were similar to women from other parts of the world, with the most abundant species being *L. iners* followed by *L. crispatus* (1-3,20) (**Fig. 2B, Table S4**). Women with high bacterial diversity were dominated by a mixture of anaerobes, including *Gardnerella, Prevotella, Sneathia, Atopobium, Dialister* and *Megasphaera* species.

**Fig. 2D** displays metabolites robustly associated with bacterial diversity in both cohorts based on the PLS loadings in **Fig 1B**. Metabolites associated with high diversity include amines, which contribute to malodor (16-18), and a number of organic acid derivatives such as 2-hydroxyisovalerate (2HV), γ-hydroxybutyrate (GHB), 2-hydroxyglutarate and 2-hydroxyisocaproate. Low diversity was characterized by elevated amino acids, including the amine precursors lysine, ornithine and tyrosine. Many of these metabolites were detected by LC-MS, and trimethylamine (high diversity) and lactate (low diversity) were detected exclusively by this method (**Table S5**). The identities of metabolites of interest were confirmed with authentic standards when available (**Fig. 2**, asterisks).

### Succinate is not associated with diversity or clinical BV

Succinate and lactate abundance are shown in panel **E** of **Fig 2**. Succinate levels, and the succinate:lactate ratio have historically been associated with BV (21-23), and succinate has been postulated to play an immunomodulatory role (23). Here we show that succinate is not associated with bacterial diversity, nor is it significantly elevated in clinical BV as defined by Nugent scoring. This trend was independent of the detection method used. In addition, succinate was elevated in women dominated by *L. crispatus* compared with *L. iners* (unpaired ttest, Benjamini-Hochberg $p < 0.01$) (**Fig S3**), indicating *L. crispatus* may produce succinate *in vivo*, a phenomenon that has been demonstrated *in vitro* (24).

### Metabolites associated with diversity are sensitive and specific for clinical BV

We defined clinical BV by the Nugent method, which is the current gold standard for BV diagnosis. This microscopy-based technique defines BV as a score of 7-10 when low numbers of lactobacilli morphotypes are observed, and high numbers of short rods presumed to represent BV associated bacteria are present. Nugent Normal (N) is defined as a score of 1-3, indicating almost exclusively *Lactobacillus* morphotypes. Intermediate samples are given a score of 4-6 and do not fit into either group. Although Nugent scores correlated well with bacterial diversity in our study, it was apparent from the microbiota and metabolome profiles that two samples (41 and 145) had been misclassified by Nugent (**Fig. 2A,** red dots). The Nugent status of these samples was therefore corrected prior to all further analyses.

In total we identified 49 metabolites that were significantly different between BV and N (unpaired t-test, Benjamini-Hochberg $p < 0.01$, **Table S2**). Nineteen of these have not been reported as differential in the literature, and 12 could not be identified. We determined the odds ratio (OR) for BV based on conditional logistic regressions of all individual metabolites detected by GC-MS (**Table S2**) to determine if the metabolites we associated with high bacterial diversity could accurately identify clinical BV as defined by Nugent scoring. Metabolites significantly elevated in Nugent BV (unpaired t-test, Benjamini-Hochberg $p < 0.01$) with OR > 1 are shown in **Fig. 3A**. Succinate was included as a comparator, although it did not reach significance. Both GHB and 2HV were significantly higher in women with BV, and had OR > 2.0, demonstrating they are novel indicators not only of high bacterial diversity, but also clinical BV. Receiver operating characteristics (ROC) curves built from LC-MS data determined that high 2HV, high GHB, low lactate and low tyrosine were the most sensitive and specific biomarkers for BV, with the largest area under the curve (AUC) achieved using the ratio of 2HV:tyrosine (AUC=0.993)(**Fig 3B-D**). ROC curves of GC-MS data identified similar trends, with the largest AUC achieved by the ratio of GHB:tyrosine (AUC=0.968) (**Table S6**).

We determined the optimal cut points for the GHB:tyrosine (0.621) and 2HV:tyrosine (0.882) ratios by selecting values which maximized the sensitivity and specificity for BV. Nugent intermediate samples grouped equally with N or BV based on these cut points, and intermediate-scored samples with smaller proportions of lactobacilli tended to group with BV (**Fig 4**).

## Validation of biomarkers in a blinded replication cohort from Tanzania

We validated these biomarkers in a blinded cohort of 45 pregnant women from Mwanza, Tanzania (Bisanz at al, manuscript submitted). Using the 2HV:tyrosine cut point identified in the Rwanda data set, we identified Nugent BV with 89% sensitivity and 94% specificity in the validation set (AUC=0.946), demonstrating our findings are reproducible in an ethnically distinct population (**Fig. 5, Table S7**). The GHB:tyrosine ratio cut point was slightly less specific (88%), with an AUC of 0.948. We confirmed that succinate was not significantly different between Nugent N and BV in the validation set, nor was the succinate:lactate ratio.

## Identification of *G. vaginalis* as a producer of GHB

Correlations between metabolites and all taxa indicated that tyramine, putrescine, and cadaverine were most correlated with *Dialister* (Pearson's R = 0.53, 0.58, 0.69, $p < 0.01$) (**Table S8**), indicating this genus may contribute to malodor. We found that GHB was most correlated with *G. vaginalis* (Pearson's R = 0.66, $p< 0.01$), while 2HV was most correlated with *Dialister*, *Prevotella*, and *Atopobium* (Pearson's R = 0.61, 0.58, 0.55, $p < 0.01$).

We chose to investigate the correlation between GHB and *G. vaginalis*, since this metabolite was novel, and predictive for both Shannon's diversity and Nugent BV. Examination of available genomes showed that many strains of *G. vaginalis* possess a putative GHB dehydrogenase (annotated as 4-hydroxybutyrate dehydrogenase). We extracted metabolites from bacterial colonies grown on agar plates and reproducibly detected GHB in *G. vaginalis* extracts well above control levels (unpaired t-test, $p < 0.05$), but did not detect GHB from other species commonly associated with BV (**Fig. 6, Table S9**). These data suggest that *G. vaginalis* is the primary source of GHB detected *in vivo*.

## Discussion

We have demonstrated that the vaginal metabolome is strongly correlated with bacterial diversity in both pregnant and non-pregnant Rwandan women, and identified 2HV and GHB as novel biomarkers of clinical BV, the latter of which we attribute to production by *G. vaginalis*. We obtained extremely accurate results by controlling for the mass of vaginal fluid collected, however we recognize this may not be logistically possible in a clinical setting. To circumvent this need we expressed biomarkers as ratios to the amino acid tyrosine, the most differential amino acid in health. Given the highly conserved nature of the vaginal microbiota across different populations and ethnicities (1-3,20), we expect these biomarkers to be globally applicable for the diagnosis of BV, and our ability to replicate findings in a distinct population strongly supports this theory.

Although *G. vaginalis* is strongly correlated with GHB in the vagina, it is important to note that no single organism has been identified as the cause of BV, and *G. vaginalis* is present in many women with a lactobacilli-dominated microbiota. However, as GHB is metabolized from succinate in other bacteria (25,26), a similar pathway could exist in *G. vaginalis*. Succinate-producing genera may therefore be required, making *G. vaginalis* essential, but not sufficient for GHB production in the vagina. This remains to be tested.

The finding that succinate, an end product of anaerobic respiration, was not significantly elevated in women with BV was an unexpected outcome. This metabolite has historically been associated with the condition, but has not been tested in the context of a large untargeted metabolomic study. Other groups have reported large ranges in succinate abundance in women with BV (21,22), or used pooled samples (22). This, and our employment of necessary multiple testing corrections, could account for disparities in results. Differences in succinate abundance may have been more pronounced in previous studies if there were a lack of *L. crispatus* dominated women, which our data indicates is a succinate producer. There is increased expression of succinate producing pathways during BV (27), and therefore it is probable that large amounts are produced initially, but then rapidly converted to other compounds, such as GHB, by the microbiota and/or host.

In addition to GHB, 2HV was identified as a highly specific novel biomarker for BV. 2HV is produced from breakdown of branched chain amino acids in humans (28) and some bacteria (29-31). When the trend for amino acid depletion in BV is considered, these findings suggest increased amino acid catabolism in this condition. Some of these amino acids are converted to the amines cadaverine, tyramine, and putrescine, which are also associated with BV. These odor-causing compounds were most correlated with *Dialister*. Yeoman *et al.* (32) also linked amines to *Dialister* species, and the decarboxylating genes required for amine production are expressed by this genus *in vivo* (27). These data strongly suggest that *Dialister* is one of the genera responsible for malodor in the vagina. Given the small proportion of this genus in women with BV (0.2-8% in our study), this emphasizes the need for functional characterizations of the microbiome using metabolomic and transcriptomic approaches.

The exact role, if any, of GHB and 2HV in the etiology of BV is unknown. Systemically GHB has both inhibitory and excitatory effects through activation of the GABA(B) and perhaps GABA(A) receptors in the brain, resulting in stimulatory and sedative effects if taken at high doses (33-35). The effects of GHB at other sites remain elusive. Future work should attempt to elucidate biological function of GHB and other novel metabolites to determine what effect (if any) they have on lactobacilli and the vaginal environment.

In summary, we have demonstrated using an untargeted, multiplatform approach that differences in the vaginal metabolome are driven by bacterial diversity. Other metabolomic studies have focused on symptom-associated metabolites (31), changes after treatment (36), or longitudinal changes in a few subjects (37), and included exclusively non-pregnant women. We identified several novel biomarkers for clinical BV that are independent of pregnancy status, and replicated this result in a blinded cohort.

By combining high-throughput sequencing with advanced mass spectrometry techniques we have shown how *in vivo* metabolite information can be used to identify sources of metabolic end products in bacterial communities. These techniques can be applied to many systems where organisms may be fastidious or unculturable, and provide a much-needed link between microbial composition and function.

## Methods

### Clinical samples

Premenopausal women between the ages of 18 and 55 were recruited at the University of Kigali Teaching Hospital (CHUK) and the Nyamata District Hospital in Rwanda. The Health Sciences Research Ethics Board at Western University, Canada, and the CHUK Ethics Committee, Rwanda granted ethical approval for the study. Participants were excluded if they had reached menopause, had a current infection of gonorrhoea, Chlamydia, genital warts, active genital herpes lesions, active syphilis, urinary tract infections, received drug therapy that may affect the vaginal microbiome, had unprotected sexual intercourse within the past 48 hours, used a vaginal douche, genital deodorant or genital wipe in past 48 hours, had taken any probiotic supplement in past 48 hours, or were menstruating at time of clinical visit. After reviewing details of the study, participants gave their signed consent before the start of the study. For metabolome analysis, sterile Dacron polyester-tipped swabs (BD) were pre-cut with sterilized scissors and weighed in 1.5 ml microcentrifuge tubes prior to sample collection.

Using sterile forceps to clasp the pre-cut swabs, a nurse obtained vaginal samples for metabolomic analysis by rolling the swab against the mid-vaginal wall. A second full-length swab was obtained for Nugent Scoring and 16S rRNA gene sequencing using the same method. Nugent Scoring was performed at CHUK by Amy McMillan. Vaginal pH was measured using pH strips. Samples were frozen within 2 hours of collection and stored at -20 °C or below until analysis.

**Microbiome profiling**

Vaginal swabs for microbiome analysis were extracted using the QIAamp DNA stool mini kit (Qiagen) with the following modifications: swabs were vortexed in 1 mL buffer ASL before removal of the swab and addition of 200 mg of 0.1 mm zirconia/silica beads (Biospec Products). Samples were mixed vigorously for 2 x 30 seconds at full speed with cooling at room temperature between (Mini-BeadBeater; Biospec Products). After heating to 95 °C for 5 minutes, 1.2 ml of supernatant was aliquoted into a 2ml tube and one-half an inhibitEx tablet (Qiagen) was added to each sample. All other steps were performed as per the manufacturers instructions. Sample amplification for sequencing was carried out using the forward primer (ACACTCTTTCCCTACACGACGCTCTTCCGATCTnnnn(8)CWACGCGARGAACCTTACC) and the reverse primer (CGGTCTCGGCATTCCTGCTGAACCGCTCTTCCGATCTn(12)ACRACACGAGCTGACGAC) where nnnn indicates four randomly incorporated nucleotides, and (8) was a sample nucleotide specific barcode. The 5' end is the adapter sequence for the Illumina MiSeq sequencer and the sequences following the barcode are complementary to the V6 rRNA gene region. Amplification was carried out in 42 µL with each primer present at 0.8 pMol/mL, 20 µL GoTaq hot start colorless master mix (Promega) and 2 µL extracted DNA. The PCR protocol was as follows: initial activation step at 95 °C for 2 minutes and 25 cycles of 1 minute 95 °C, 1 minute 55 °C and 1 minute 72 °C.

All subsequent work was carried out at the London Regional Genomics Centre (LRGC, lrgc.ca, London, Ontario, Canada). Briefly, PCR products were quantified with a Qubit 2.0 Flourometer and the high sensitivity dsDNA specific fluorescent probes (Life Technologies). Samples were mixed at equimolar concentrations and purified with the QIAquick PCR Purification kit (QIAGEN). Samples were paired-end sequenced on an Illumina Mi-Seq with the 600 cycle version 3 reagents with 2x220 cycles. Data was extracted from only the first read, since it spanned the entirety of the V6 region including the reverse primer and barcode.

Resulting Reads were extracted and de-multiplexed using modifications of in-house Perl and UNIX-shell scripts with operational taxonomic units (OTUs) clustered at 97% identity, similar to our reported protocol (38). Automated taxonomic assignments were carried out by examining best hits from comparison the Ribosomal Database Project (rdp.cme.msu.edu) and manually curated by comparison to the Green genes database (greengenes.lbl.gov) and an in house database of vaginal sequences (Macklaim unpublished). Taxa with matches at least 95% similarity to query sequences were annotated as such. OTUs were summed to the genus level except for lactobacilli, and rare OTUs found at less than 0.5% abundance in any sample removed. **Table S1** displays the nucleotide barcodes and their corresponding samples. Reads were deposited to the Short Read Archive (BioProject ID: xxx). To control for background contaminating sequences, a no-template control was also sequenced. Barplots were constructed with R {r-project.org } using proportional values.

To avoid inappropriate statistical inferences made from compositional data, centred log-ratios (clr), a method previously described by Aitchison (39) and adapted to microbiome data was used with paired t-tests for comparisons of genus and species level data (40). The Benjamini Hochberg (False Discovery rate) method was used to control for multiple testing with a significance threshold of 0.1. All statistical analysis, unless otherwise indicated, was carried out using R (r-project.org).

**Sample Preparation GC-MS**

Vaginal swabs were pre-cut into 1.5 mL tubes and weighed prior to and after sample collection to determine the mass of vaginal fluid collected. After thawing, swabs were eluted in methanol-water (1:1) in 1.5 mL microcentrifuge tubes to a final concentration of 50 mg vaginal fluid/mL, which corresponded to a volume ranging from 200-2696 µL, depending on the mass of vaginal fluid collected. A blank swab eluted in 800 µL methanol-water was included as a negative control. All samples were vortexed for 10 s to extract metabolites, centrifuged for 5 min at 10 621 g, vortexed again for 10 s after which time the brushes were removed from tubes. Samples were centrifuged a final time for 10 min at 10 621 g to pellet cells and 200 µL of the supernatant was transferred to a GC-MS vial. The remaining supernatant was stored at -80 °C for LC-MS analysis. Next, 2 µL of 1 mg/mL ribitol was added to each vial as an internal standard. Samples were then dried to completeness using a SpeedVac. After drying, 100 µL of 2% methoxyamine-HCl in pyridine (MOX) was added to each vial for derivatization and incubated at 50 °C for 90 min. 100 µL N- Methyl-N-(trimethylsilyl) trifluoroacetamide (MSTFA) was then added and incubated at 50 °C for 30 min. Samples were then transferred to micro inserts before analysis by GC-MS (Agilent 7890A GC, 5975 inert MSD with triple axis detector). 1 µL of sample was injected using pulsed splitless mode into a 30 m DB5-MS column with 10 m duraguard, diameter 0.35mm, thickness 0.25 µm (JNW Scientific). Helium was used as the carrier gas at a constant flow rate of 1 ml/min. Oven temperature was held at 70 °C for 5 min then increased at a rate of 5 °C/min to 300 °C and held for 10 min. Solvent delay was set to 13 min to avoid solvent and a large lactate peak, and total run time was 61 min. Masses between 25 m/z and 600 m/z were selected by the detector. All samples were run in random order and a standard mix containing metabolites expected in samples was run multiple times throughout to ensure machine consistency.

**Data Processing GC-MS**

Chromatogram files were de-convoluted and converted to ELU format using the AMDIS Mass Spectrometry software (41), with the resolution set to high and sensitivity to medium. Chromatograms were then aligned and integrated using Spectconnect (42) (http://spectconnect.mit.edu), with the support threshold set to low. All metabolites found in the blank swab, or believed to have originated from derivatization reagents were removed from analysis at this time. After removal of swab metabolites, the IS matrix from Spectconnect was transformed using the additive log ratio transformation (alr) (39) and ribitol as a normalizing agent (log2(x) / log2(ribitol)). Zeros were replaced with two thirds the minimum detected value on a per metabolite basis prior to transformation. All further metabolite analysis was performed using these alr transformed values.

Metabolites were initially identified by comparison to the NIST 11 standard reference database (http://www.nist.gov/srd/nist1a.cfm). Identities of metabolites of interest were then confirmed by authentic standards if available.

**Whole metabolome analysis**

In order to visualize trends in the metabolome as detected by GC-MS, principal component analysis (PCA) was performed using pareto scaling. To determine the percentage of variation in the metabolome that could be explained by a single variable we performed a series of partial least squares (PLS) regressions where each variable was used as a continuous latent variable. We tested every taxa, pH, Nugent score, pregnancy status, Shannon's diversity index and sample ID and compared the percent variation explained by the first component of each PLS. The variable with the highest value was determined to be most closely associated with the metabolome (Shannon's Diversity). Analysis was conducted in R using the PLS package and unit variance scaling. Jackknifing with 20% sample removal and 10 000 repetitions was then applied to determine 95% confidence intervals for each metabolite. Metabolites with confidence intervals that did not cross zero in both cohorts (pregnant and non-pregnant) were considered significantly associated with diversity. Heatmaps of significant metabolites were constructed using the heatmap.2 function in R with average linkage hierarchical clustering and manhattan distances. Unless specified otherwise, all tests for differential metabolites between groups were performed using unpaired t-test with a Benjamini-Hochberg (False Discovery Rate) significance threshold of $p < 0.01$ to account for multiple testing and multiple group comparisons. Correlations between metabolites and taxa were performed using alr transformed values for metabolites and clr values with 128 Monte Carlo instances for microbiota data in R using the ALDEx2 package (40).

Odds ratios of metabolites to identify Nugent BV from Normal were calculated from conditional logistic regressions performed on all metabolites using the glm function in R with 10 000 iterations and a binomial distribution. Metabolites with 95 % CI > 1 and $p < 0.01$ (unpaired t-test, Benjamini-Hochberg corrected) were determined to be significantly elevated in Nugent BV. "Nugent BV" was defined by the clinical definition of a score of 7-10, with a score of 0-3 being "Nugent Normal". ROC curves and forest plots were built in R using the pROC and Gmisc packages respectively.

**Sample Preparation LC-MS**

To confirm GC-MS findings, samples which had at least 100 µL remaining after GC-MS were also analyzed by LC-MS. 100 µL of supernatant was transferred to vials with microinserts and directly injected into an Agilent 1290 Infinity HPLC coupled to a Q-Exactive mass spectrometer (Thermo Fisher Scientific) with a HESI source. For HPLC, 2 µL of each sample was injected into a ZORBAX Eclipse plus C18 2.1 x 50mm x 1.6 micron column. Mobile phase (A) consisted of 0.1% formic acid in water and mobile phase (B) consisted of 0.1% formic acid in acetonitrile. The initial composition of 100% (A) was held constant for 30 s and decreased to 0% over 3.0 min. Mobile phase A was then held at 0% for 1.5 minutes and returned to 100% over 30s for a total run time of 5 min.

Full MS scanning between the ranges of m/z 50-750 was performed on all samples in both positive in negative mode at 140 000 resolution. The HESI source was operated under the following conditions: nitrogen flow of 25 and 15 arbitrary units for the sheath and auxiliary gas respectively, probe temperature and capillary temperature of 425 °C and 260 °C respectively and spray voltage of 4.8 kV and 3.9 kV in positive and negative mode respectively. The AGC target and maximum injection time were 3e6 and 500 ms respectively. For molecular characterization, every tenth sample was also analyzed with a data dependent $MS^2$ method where a 35 000 resolution full MS scan identified the top 10 signals above a 8.3e4 threshold which were subsequently selected at a 1.2 m/z isolation window for $MS^2$. Collision energy for $MS^2$ was 24, resolution 17 500, AGC target 1e5 and maximum injection time was 60ms. Blanks of pure methanol were run between every sample to limit carryover, and a single sample was run multiple times with every batch to account for any machine inconsistency. A blank swab extract was also run as a negative control.

For increased sensitivity, a separate LC-MS method was used for relative quantification of GHB in human samples. This was accomplished by selected ion monitoring in the mass range of 103.1 – 107.1 m/z in positive mode, and integrating the LC peak area of the $[M+H^+]$ ion (± 5 ppm).

**Data Processing LC-MS**

After data acquisition Thermo .RAW files were converted to .MZML format using

ProteoWizard (43) and imported into MZmine 2.11 (44) (http://mzmine.sourceforge.net) for chromatogram alignment and deconvolution. Masses were detected using the Exact Mass setting and a threshold of 1E5. For Chromatogram Builder, minimum time was 0.05 min, minimum height 3E3, and m/z threshold set to 0.025 m/z or 8 ppm. Chromatogram Deconvolution was achieved using the Noise Amplitude setting with the noise set to 5E4 and signal to 1E5 for negative mode. Due to an overall greater signal and noise in positive mode, the noise was adjusted to 6E5 and signal to 6.5E5 for positive mode. Join aligner was used to combine deconvoluted chromatograms into a single file with the m/z threshold set to 0.05 m/z or 10 ppm, weight for m/z and RT set to 20 and 10 respectively, and a RT tolerance of 0.4 min. After chromatograms were aligned, a single .CSV file was exported and all further analysis was carried out in R.

To confirm metabolites identified as significant by GC-MS in the LC-MS data set, the masses of metabolites of interest were searched in the LC-MS data set, and identities confirmed by $MS^2$ using METLIN (45) and the Human Metabolome Database (46) online resources. Standards of metabolites of interest were also run to confirm identities when available. An unpaired t-test with Benjamini-Hochberg correction was used to determine metabolites significantly different between Nugent BV and Normal in the LC-MS data set. Metabolites with corrected $p < 0.05$ were considered statistically significant. Metabolites detected exclusively by LC-MS that have previously been associated with BV or health (lactate, trimethylamine) were also included in this analysis. Data was log base 10 transformed prior to data analysis and zeros replaced by 2/3 the minimum detected value on a per metabolite basis. To determine optimal cut points of biomarkers for diagnostic purposes, cut points were computed from LC-MS data using the OptimalCutpoints package in R (47) and the Youden Index method (48).

### Validation in blinded replication cohort

Women between the ages of 18 and 40 were recruited from an antenatal clinic at the Nyerere Dispensary in Mwanza, Tanzania as part of a larger study on the effect of micronutrient supplemented probiotic yogurt on pregnancy. The study was approved by both the Medical research Coordinating Committee of the National Institute for Medical Research (NIMR), as well as from the Health Sciences Research Ethics Board at Western University. The study was registered with clinicaltrials.gov (NCT02021799). Samples were collected using the methods mentioned above, and Nugent scores performed by research technicians at NIMR in Mwanza, Tanzania. A subset of samples was selected based on these Nugent scores by a third party, who ensured there was not repeated sampling of any women. Amy McMillan, who performed metabolite analysis, was blinded to the Nugent scores for the duration of sample processing and data analysis. Biomarkers were quantified in samples by LC-MS using the protocols mentioned above. The study was unblinded after the submission of BV status based on the ratio cut points established in the Rwandan data set.

### Identification of putative GHB dehydrogenases in *G. vaginalis* strains

The protein sequence of a *bona fide* 4-hydroxybutyrate (GHB) dehydrogenase isolated from Clostridium kluyveri (25) (GI:347073) was blasted against all strains of *G. vaginalis* in the NCBI protein database. Blast results identified multiple isolates containing a putative protein with 44-46% identity to the GHB dehydrogenase from *C. kluyveri*. The strain used for *in vitro* experiments (*G.vaginalis* ATCC 14018) was not present in the NCBI protein database, however a nucleotide sequence in 14018 with 100% nucleotide identity to a putative 4-hydroxybutyrate dehydrogenases in strain ATCC 14019 (GI:311114893) was identified, indicating potential for GHB production by strain 14018.

### *In vitro* extraction of GHB from vaginal isolates

Due to their fastidious nature, we found it difficult to obtain consistent growth of all vaginal strains in liquid media. To circumvent this, a lawn of bacteria was plated and metabolites were extracted from agar punches. All strains were grown on Columbia Blood Agar (CBA) plates using 5% sheep's blood for 96h under strict anaerobic conditions, with the exception of *L. crispatus*, which was grown on de Man Rogosa Sharp (MRS) agar for 48 h. To extract metabolites, 16 agar punches 5 mm in diameter were taken from each plate and suspended in 3 mL 1:1 Me:$H_2$0. Samples were then sonicated in a water bath sonicater for 1h, transferred to 1.5 ml tubes after vortexing and spun in a desktop microcentrifuge for 10 min at 10 621 g to pellet cells. 200 µl of supernatant was then aliquoted for GC-MS described above. The area of each peak was integrated using ChemStation (Agilent) by selecting m/z 233 in the range of 14-16 min. Initial peak width was set to 0.042 and initial threshold at 10. An authentic standard of GHB was run with samples to confirm identification. Un-inoculated media was used as a control and experiments were repeated three times with technical duplicates.

## Acknowledgments

We would like to thank the staff at CHUK and Nyamata hospital for their participation in patient recruitment and verbal translating as well as the scientists at NIMR Mwanza who performed Nugent scoring for the replication cohort. We thank Tim McDowell for technical assistance with metabolomics platforms. Also to Jeremy Burton for assistance with logistics of sample collection and helpful discussion, and Megan Enos and Shannon Seney for the Tanzanian cohort study. This work was partially funded by a SFA grant from CIDA and by a Team Grant from CIHR to the Vogue Research Group. Funding sources to individuals: AM by CIHR and Ontario Graduate Scholarship.

## Author Contributions

A.M. participated in study design, supervised patient recruitment and sample collection, performed sample processing, participated in method development for metabolite profiling, analyzed and interpreted microbiome and metabolome data, performed *in vitro* experiments and wrote the manuscript. S.R. participated in study design and co-ordinated patient recruitment and ethics approval. M.S. provided platforms for metabolite profiling, advised on method development and analysis for metabolomics, and contributed to manuscript generation. J.M. participated in study design, advised on data analysis and contributed to manuscript generation. J.R. participated in LC-MS method development and data analysis. J.B. participated in acquisition of Tanzanian replication cohort and blinded samples. G.B.B. developed methods for integration of microbiome and metabolome data, supervised all data analysis and contributed to manuscript generation. G.R. conceived and designed the study, acquired funding and contributed to manuscript generation.

## Figures

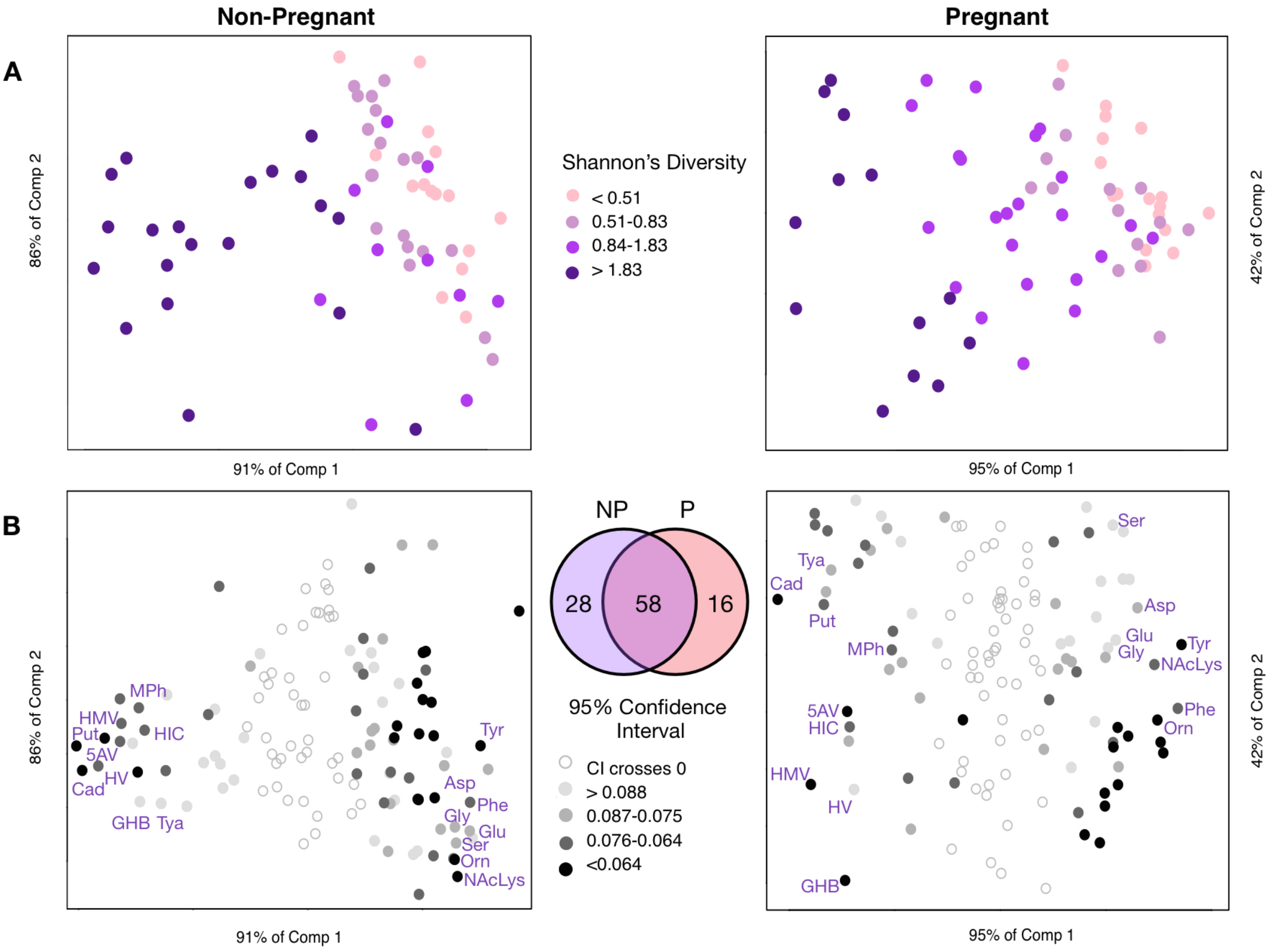

**Fig. 1.** The vaginal metabolome is most correlated with bacterial diversity. All analyses were carried out independently for non-pregnant (left) and pregnant (right) cohorts. Row (**A**) Partial least squares regression (PLS) scoreplot built from 128 metabolites detected by GC-MS using bacterial diversity as a continuous latent variable. Each point represents a single woman (n=131). The position of points display similarities in the metabolome, with samples closest to one another being most similar. Circles are colored by diversity of the microbiota measured using Shannon's diversity, where darker circles indicate higher diversity. Row (**B**) PLS regression loadings. Each point represents a single metabolite. Shaded circles indicate metabolites robustly associated with diversity in either cohort (Jackknifing, 95% CI < 0 > ). Shading of circles corresponds to the size of the confidence interval (CI) for each metabolite, where darker circles indicate narrower CIs. Venn diagram depicts overlap between metabolites associated with diversity in either cohort. Cad:Cadaverine, Tya:Tyramine, Put:Putrescine, MPh:Methylphosphate, 5AV:5-aminovalerate, HIC:2-hydroxyisocaproate, HMV:2-hydroxy-3-methylvalerate, HV:2-hydroxyisovalerate, GHB: γ-hydroxybutyrate. Ser:serine, Asp:aspartate, Glu:glutamate, Gly:glycine, Tyr:tyrosine. NAcLys:n-acetyl-lysine, Phe:phenylalanine, Orn:ornithine.

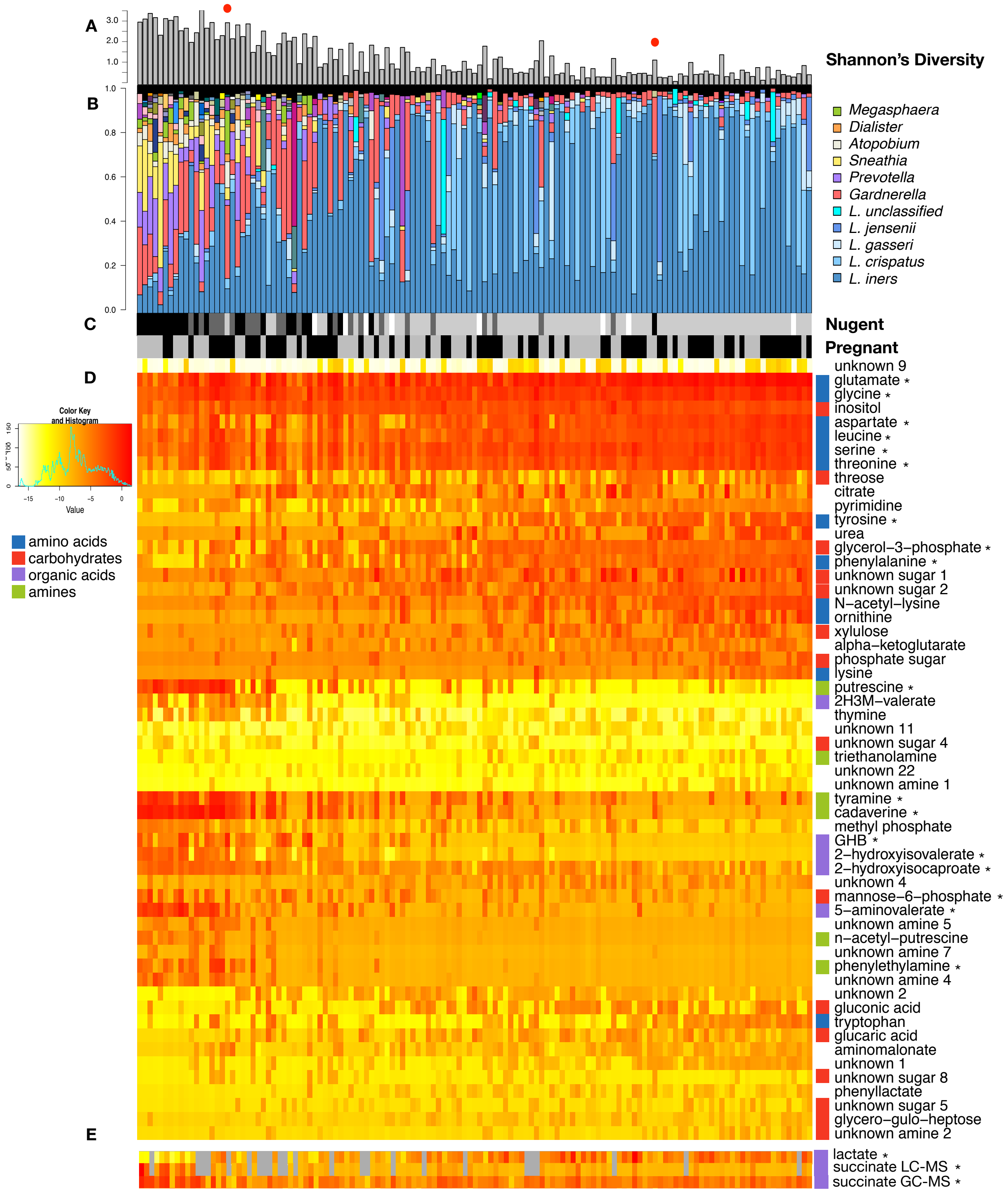

**Fig. 2.** Bacterial taxa and metabolites correlated with bacterial diversity in the vagina. Cohorts (non-pregnant and pregnant) were combined prior to analyses. Samples are ordered by their position on the first component (x-axis) of a partial least squares regression (PLS) built from metabolites using bacterial diversity as a continuous latent variable (see Fig. S2). Diversity was calculated using Shannon's diversity (**A**). Red dots indicate samples clearly misclassified by Nugent. Barplots (**B**) display the vaginal microbiota profiled using the V6 region of the 16S rRNA gene. Each bar represents a single sample from a single woman, and each colour a different bacterial taxa. (**C**) Nugent Score (black=7-10 (BV), dark grey=4-6 (Int), light grey=1-3 (N), white=ND) and pregnancy status (black=P, grey=NP). (**D**) Heatmap of GC-MS detected metabolites which were robustly associated with diversity in both cohorts (Jackknifing, 95% CI <0>). Metabolites are clustered using average linkage hierarchical clustering. (**E**) Lactate and succinate abundance. Grey = ND. (*) indicates metabolites confirmed by authentic standards.

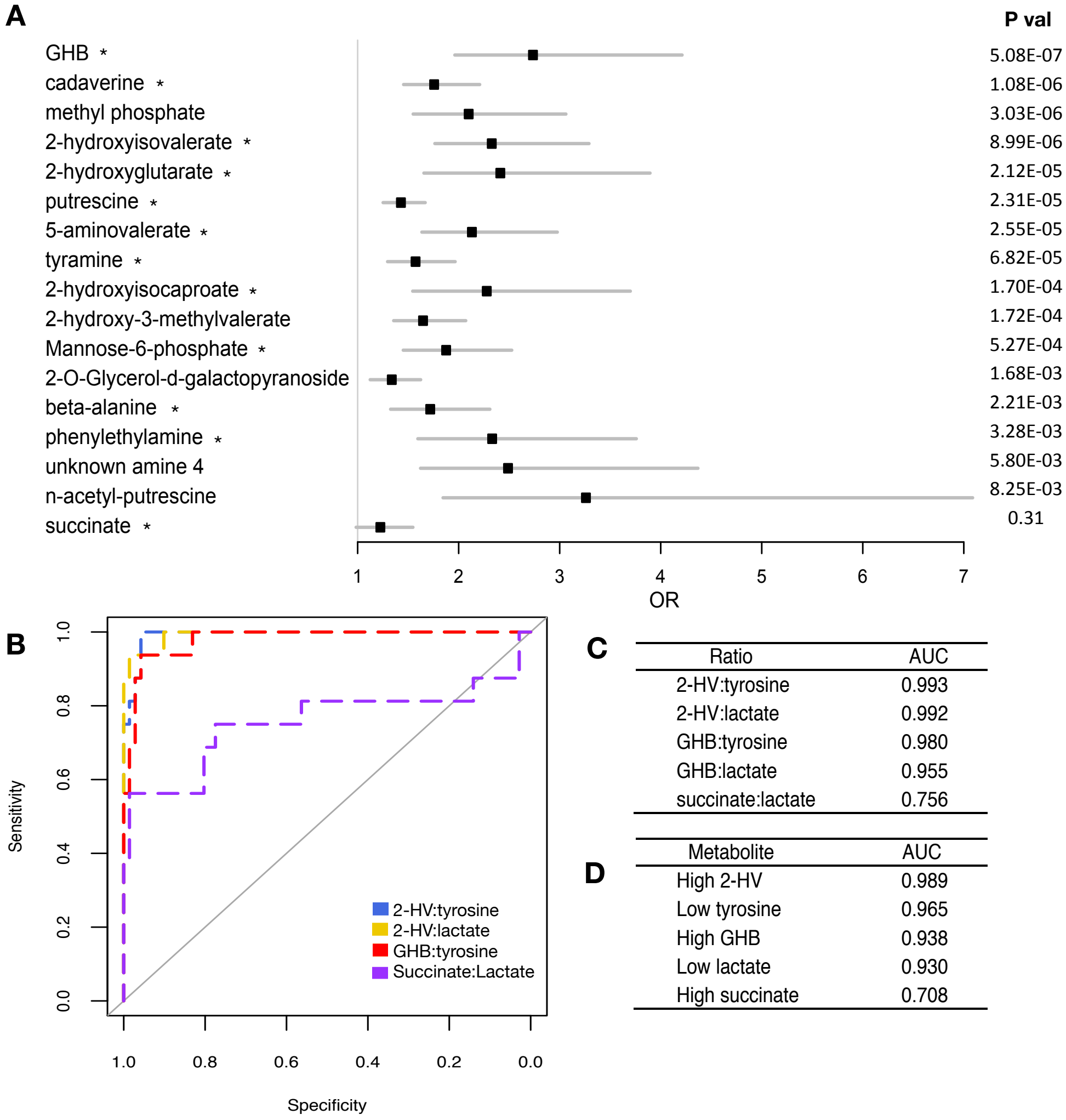

C

| Ratio | AUC |
|---|---|
| 2-HV:tyrosine | 0.993 |
| 2-HV:lactate | 0.992 |
| GHB:tyrosine | 0.980 |
| GHB:lactate | 0.955 |
| succinate:lactate | 0.756 |

D

| Metabolite | AUC |
|---|---|
| High 2-HV | 0.989 |
| Low tyrosine | 0.965 |
| High GHB | 0.938 |
| Low lactate | 0.930 |
| High succinate | 0.708 |

**Fig. 3.** Comparison of biomarkers to identify Nugent BV from Nugent N. (**A**) Odds ratios (OR) of metabolites with positive predictive value to identify Nugent BV. Bars represent 95% Confidence Intervals. Metabolites were detected by GC-MS and P values generated from unpaired t-tests with a Benjamini-Hochberg correction to account for multiple testing ($p < 0.01$). (*) indicates metabolites confirmed by authentic standards. (**B**) Receiver operating characteristic (ROC) curves of metabolite ratios to identify Nugent BV from Nugent N. Ratios with largest area under the curve (AUC) are shown, along with succinate:lactate as a comparator. (**C**) AUC of selected metabolite ratios to identify Nugent BV. (**D**) AUC of metabolites alone to identify Nugent BV. Panels **B**-**D** were built from LC-MS data. GHB:γ-hyroxybutyrate, 2-HV:2-hydroxyisovalerate.

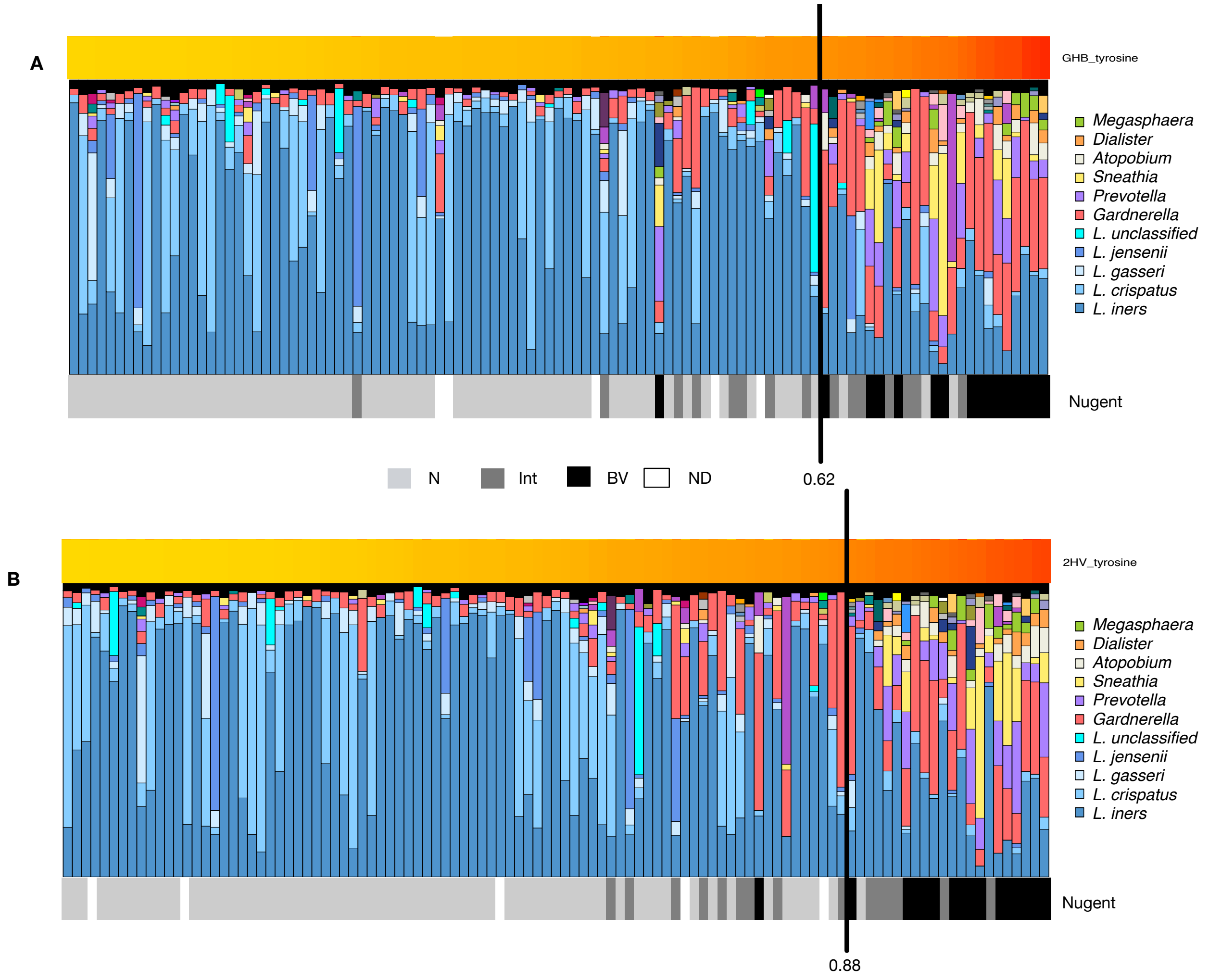

**Fig. 4**. Biomarker cut points effectively group Nugent Intermediate samples as BV or N. Barplots display the vaginal microbiota of Rwandan women sorted by (**A**) GHB:tyrosine or (**B**) 2HV:tyrosine. Each bar represents a single sample from a single woman and each colour a different bacterial taxa. Nugent scores are indicated below barplots. Black lines indicate ratio cut point for Nugent BV. Ratios were calculated from LC-MS data.

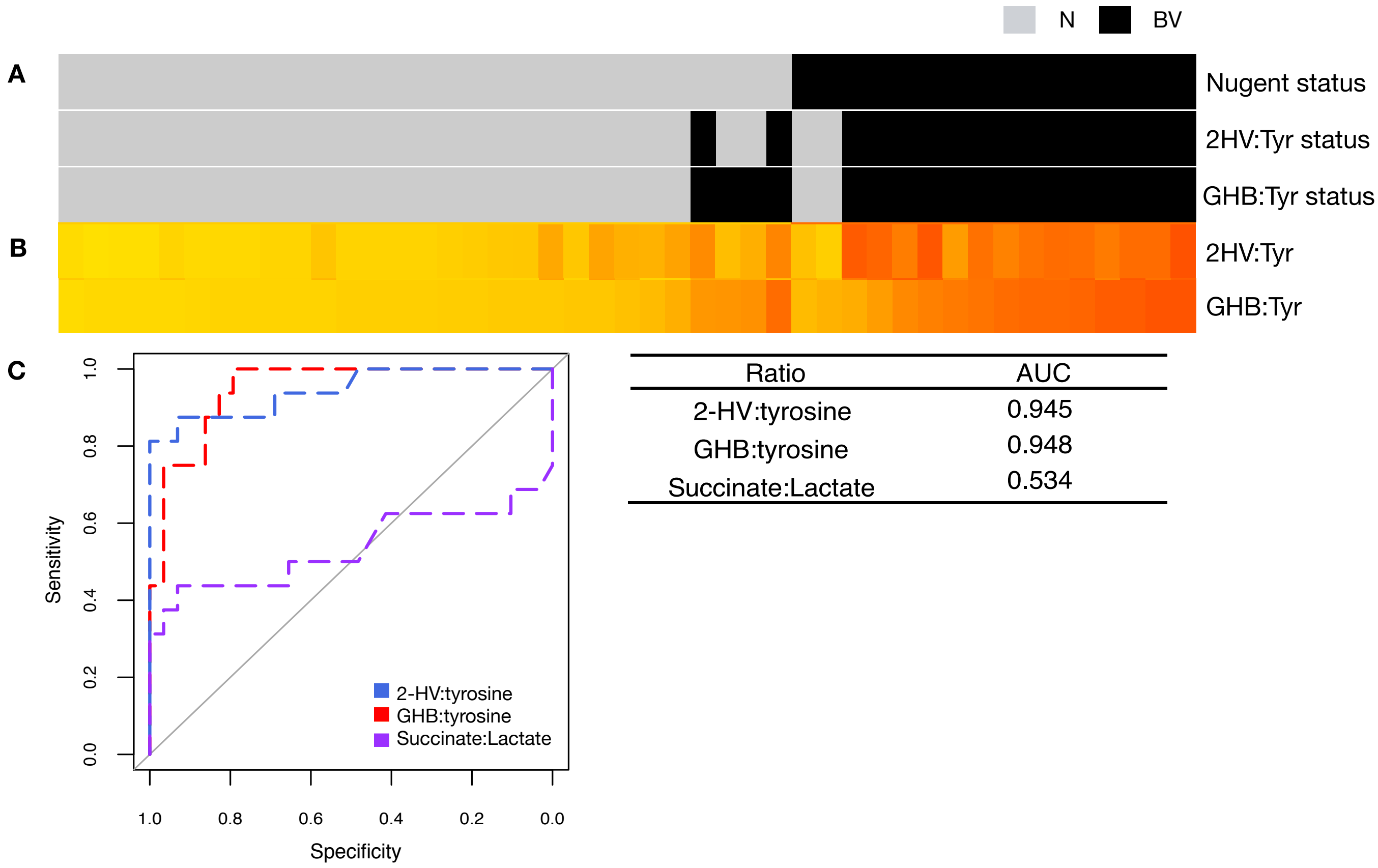

| Ratio | AUC |
|---|---|
| 2-HV:tyrosine | 0.945 |
| GHB:tyrosine | 0.948 |
| Succinate:Lactate | 0.534 |

**Fig. 5.** Biomarker validation in a blinded replication cohort of 45 women from Tanzania. (**A**) BV status as defined by Nugent Score or ratio cut points identified in the Rwandan discovery data set. Black=BV, Gray=N. (**B**) Heatmap of ratio values. (**C**) ROC curves and AUC of ratios to identify Nugent BV from N in the validation set. 2HV: 2-hydroxyisovalerate, GHB: γ-hydroxybutyrate, Tyr: tyrosine.

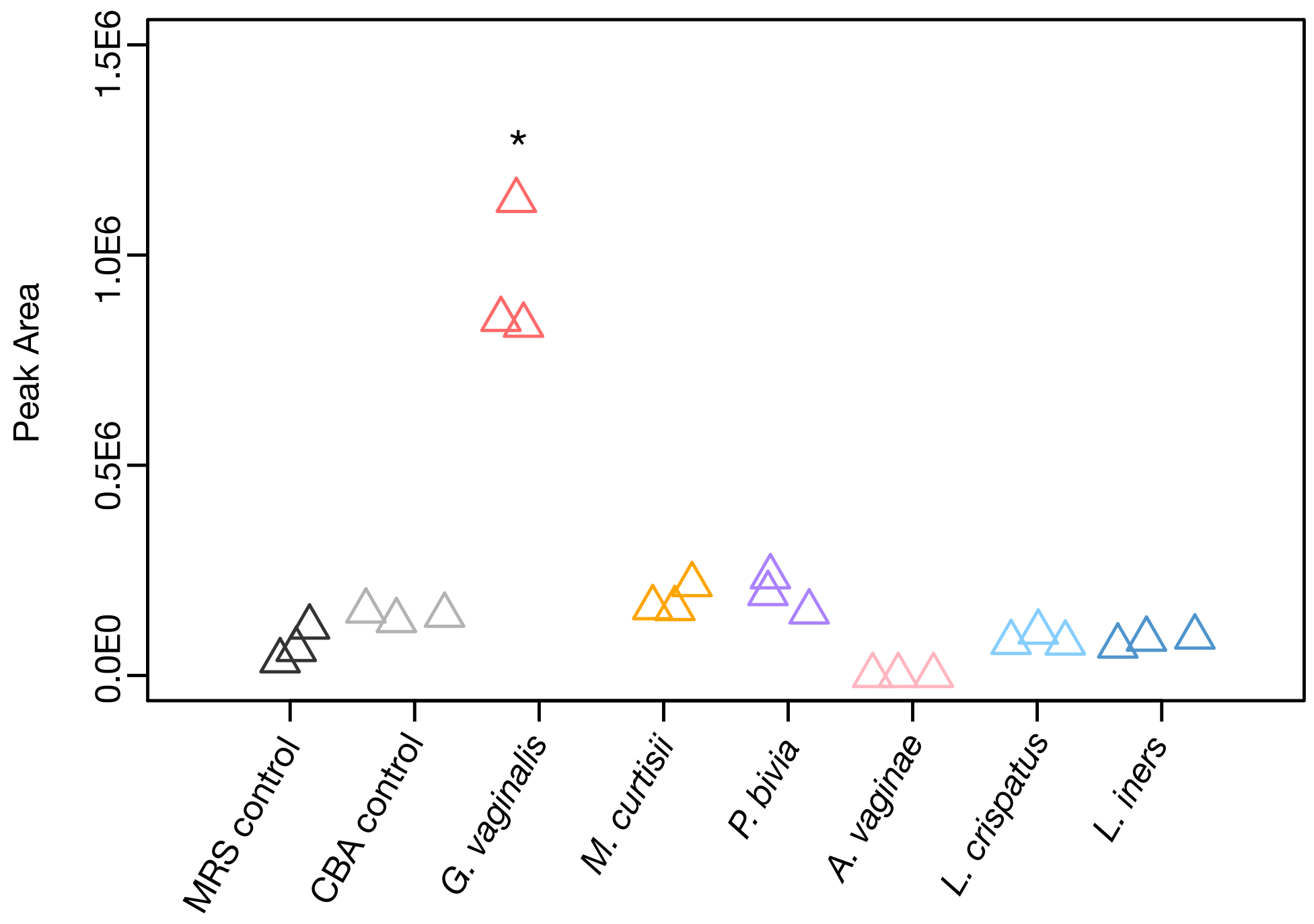

**Fig. 6**. GHB is produced by *Gardnerella vaginalis*. GHB was extracted from bacteria grown on agar plates and detected by GC-MS. Values from three independent experiments are shown where each point was generated from an average of technical duplicates. * $p < 0.05$, unpaired t-test.